\documentclass[fleqn,twoside]{article}
\usepackage{espcrc2}
\usepackage{graphicx}
\usepackage[figuresright]{rotating}

\newcommand{\AmS}{{\protect\the\textfont2
  A\kern-.1667em\lower.5ex\hbox{M}\kern-.125emS}}

\title{Physics with a very first low-energy beta-beam}

\author{Cristina Volpe \address[IPNO]{Institut de Physique Nucl\'eaire,\\
	        F-91406 Orsay cedex, France}}
       
\begin{document}

\begin{abstract}
We describe the importance of having low-energy (10-100 MeV) neutrino beams
produced through the decay of boosted radioactive ions (``beta-beams'').
We focus on the interest for neutrino-nucleus interaction
studies and their impact for astrophysics, nuclear and particle physics.
In particular, we discuss the relation to neutrinoless double-beta decay.
Finally, we mention the status as far as the feasibility of low-energy
beta-beams is concerned.
\vspace{1pc}
\end{abstract}

\maketitle

\section{INTRODUCTION}
\noindent
Nuclei are a wonderful laboratory for searches on fundamental issues,
such as the knowledge of the neutrino mass scale, or of the Majorana
versus Dirac nature of neutrinos. Nuclei can also be a beautiful tool
for the search of new physics. The original idea of ``beta-beams'',
first proposed by Zucchelli \cite{zu02}, enter in this category.
Beta-beams use the beta-decay of boosted radioactive ions to produce
well known electron (anti)neutrino beams,  while the conventional
way exploits the decay of pions and muons. This simple but intriguing
idea has opened new strategies, thanks to the future
radioactive ion beams,
at present under study, in various nuclear physics laboratories. In fact,
the planned intensities of $10^{11-13} ions/s$ can actually render
neutrino accelerator experiments using ions, feasible.

In the original paper \cite{zu02}, a new facility is described, based on
the beta-beam method, the central motivation being the search for CP
violation in the lepton sector -- the Maki-Nakagawa-Sakata-Pontecorvo
(MNSP) matrix, relating the neutrino flavor and mass basis, might indeed be complex.
With this aim the ions would be accelerated to 60-100 GeV/A
(or $\gamma=60-100$, where $\gamma$ is the Lorentz factor), requiring
accelerator infrastructure like the PS and SPS at CERN, as well as a
large storage ring pointing to an (enlarged) Fr\'ejus Underground
Laboratory, where a big detector would be located.

Very soon the interest of this new concept for
the production of low-energy  neutrino beams has been recognized \cite{vol04}.
Here the ions are boosted to a much lower $\gamma$, i.e. 5-15.
High energy scenarios have been proposed \cite{jj04} afterwards, requiring
different (or revised) accelerator infrastructures to boost the ions
at very high $\gamma$ ($\gamma>>$ 100).
(Note that for this reason the original scenario \cite{zu02} is sometimes
referred to in the literature as 
``standard'', or misleadingly ``low-energy''.) 
Detailed works exist at present both on the feasibility \cite{feas} 
as well as on the physics
potential of the standard \cite{stand} scenario,
contributing to determining the conditions for the best CP violation
sensitivity, in possible future searches.
A feasibility study is now ongoing within the Eurisol Design Study \cite{eurisol}.
Here we will focus on the physics potential of low-energy beta-beams.

\section{LOW-ENERGY BETA-BEAMS}

\subsection{Physics Motivations}
\noindent
The idea of establishing a facility producing low-energy neutrino beams,
based on beta-beams, has been proposed in \cite{vol04}. This
opens new opportunities, compared to the original
scenario. First one might
use the ion decay at rest as an intense neutrino source in order
to explore neutrino properties that are still poorly known, such as the
neutrino magnetic moment \cite{mc04}. In fact direct measurements
achieving improved limits are precious, since the observation of a
large magnetic moment points to physics beyond the Standard Model.

The interest of low-energy  beta-beams in the tens
of MeV, to perform neutrino-nucleus interaction studies,
has been discussed in \cite{vol04,ser04}. At present,
there is a limited number of
measurements available in this domain (essentially on three light nuclei), so
that theoretical predictions are of absolute necessity.
Getting accurate predictions can be a challenging task, as
the discrepancies on the
$^{12}$C \cite{c12} and $^{208}$Pb \cite{Pb} cross sections have been demonstrating.
Neutrino-nucleus applications are numerous and span from a better knowledge of
neutrino detector response using nuclei, like supernova observatories
or in oscillation experiments, to nuclear astrophysics, for the understanding
of processes like the nucleosynthesis of heavy elements.
(More information can be found e.g. in \cite{ku94,review}.)

In \cite{vol04} we have pointed out that performing
neutrino-nucleus interaction studies on various nuclei would improve
our present knowledge of the ``isospin'' and ``spin-isospin'' nuclear
response (the nuclear transitions involved in charged-current reactions are
in fact due to the isospin, like e.g. $t_{\pm}$, and spin-isospin,
like e.g. $\sigma t$, operators). A well known example
is given by the super-allowed Fermi
transitions (due to the isospin operator),
which are  essential for determining the unitarity of the
Cabibbo-Kobayashi-Maskawa (CKM) matrix, the analog of the MNSP matrix in the
quark sector. Another (less known but still intriguing) example is
furnished by the so-called Gamow-Teller transitions (these are due to the
spin-isospin operator) in mirror nuclei, which can be used to observe
second class currents, if any. These terms transform in an
opposite way under the {\it G-}parity transformation  -- the product of
charge-conjugation and of a
rotation in isospin space -- as the usual vector
and axial-vector terms \cite{vol03}, and are not present in the
Standard Model. In \cite{vol04} we have pointed out that spin-isospin and
isospin states of higher multipolarity (than those just mentioned) contribute
significantly to the neutrino-nucleus cross sections, as the energy of the
neutrino increases. Such contributions are larger when the nucleus is heavier.
 Since low-energy beta-beams have the specificity that the average
energy can be increased by increasing the Lorentz boost of the ions
(more precisely $ <E_{\nu}> \approx 2 \gamma Q_{\beta}$), they appear as an
appropriate tool for the study of these states.
Apart from their intrinsic interest, neutrino-nucleus interaction measurements
would put theoretical predictions for
the extrapolation to exotic nuclei useful for astrophysical applications.
on really solid grounds. 
They are also important for the open question of the neutrino nature.

One of the crucial issues in neutrino physics is
to know if neutrinos are Dirac or Majorana particles.  
The answer to this question can be furnished for example by the observation
of neutrinoless
double beta decay in nuclei, since this lepton violating process 
can be due to
the exchange of a Majorana neutrino. 
While the present limit is of about 0.2 eV
 \cite{ex}, future experiments aim at the challenging 
50 meV energy range. However, it
has been longly debated that the theoretical situation, as far as 
the half-life predictions are concerned, 
should be clarified: different calculations
present significant variations for the same candidate emitters.
Reducing these differences certainly represents an important theoretical
challenge for the future, and one might hope that dedicated experiments
will help making a step forward \cite{zu05}.
One way to constrain such calculations is by measuring related
processes, such as beta-decay \cite{su05}, muon capture \cite{kor03}, 
charge-exchange reactions \cite{jo}
and double-beta decay with the emission of two neutrinos \cite{vo03}
(the latter process
is allowed within the Standard Model and does not tell us anything about
the neutrino nature). Such a procedure has been used
since a long time. However each of these processes bring part of
the necessary information only.

Recently we have been showing that there is a very close connection between
neutrinoless double-beta decay and neutrino-nucleus interactions \cite{vol05}.
In fact, by rewriting the neutrino exchange potential in momentum space and
by using a multiple decomposition, the
two-body transition operators, involved in the
former, can be rewritten as a product of one-body operators, which are
essentially the same as the ones involved in neutrino-nucleus interactions.
(Note, however, that there keep being some differences
like for example short range correlations which
can play a role in the two-body process, but not in the one-body one.)
Therefore, besides the above-mentioned processes
an improved knowledge of the nuclear response through either low-energy
beta-beams or conventional sources (decay of muons at rest) could help
constraining the neutrinoless
half-life predictions as well.
Figures 1 and 2 show the contribution of different states
for two impinging neutrino energies on $^{48}$Ca taken just as an
illustrative example.
\begin{figure}[t]
\includegraphics[angle=-90.,width=6cm]{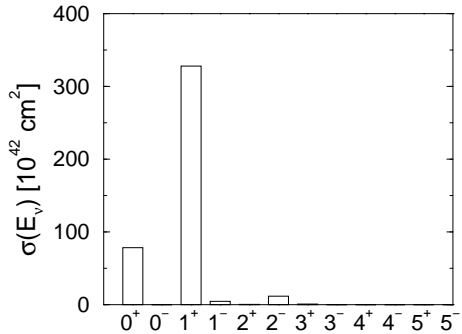}
\caption{Contribution of the states of different multipolarity
to the total charged-current  $\nu_e + ^{48}$Ca
cross section for neutrino energy $E_{\nu}=30~$MeV.
The histograms show the contribution of the Fermi
($J^{\pi}=0^+$), the Gamow-Teller  ($1^+$) and the spin-dipole
($0^-,1^-,2^-$) states and all higher multipoles up to $5$ \cite{vol05}.
\label{fig:1}}
\end{figure}
\begin{figure}[t]
\includegraphics[angle=-90.,width=6cm]{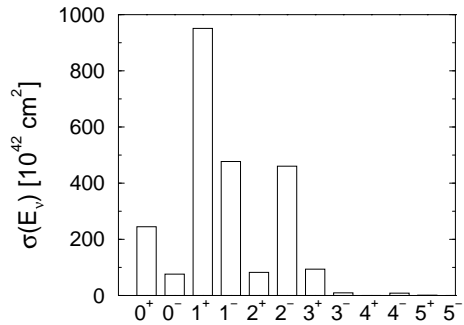}
\caption{Same as Fig.1 for $E_{\nu}=60~$MeV \cite{vol05}.  \label{fig:2}}
\end{figure}
One can see that the Gamow-Teller transition is giving the dominant
contribution at low neutrino energies, while many other states become important
when the neutrino impinging energy increases.
These states are an essential part of 
the neutrinoless double-beta decay half-lives as well \cite{vol04}.

\subsection{A small storage ring}
\noindent
The main aim of the work in Ref.\cite{ser04} has been to calculate exactly:
{\it i)} the neutrino-nucleus interactions rates expected at a 
low-energy beta-beam facility, by using parameters from the 
first feasibility study \cite{feas}; {\it ii)} to 
study how these scale by changing the geometry of the storage ring.
In particular, two sizes have been considered: a small one, i.e.
150 m straight sections and 450 total length, like the one planned for
the future GSI facility \cite{gsi}; a large one, having 2.5 km straight
sections and 7 km total length, such as the one considered in the original
beta-beam baseline scenario \cite{zu02}.
Table I shows the events for deuteron, oxygen, iron and lead,
taken as typical examples, the detector being located at 
10 m from the storage ring.
\begin{table*}
\caption{}
\label{table:1}
\newcommand{\m}{\hphantom{$-$}}
\newcommand{\cc}[1]{\multicolumn{1}{c}{#1}}
\renewcommand{\tabcolsep}{2pc} 
\renewcommand{\arraystretch}{1.2} 
\begin{tabular}{@{}lllll}
\hline
 Reaction          &  Ref.            & Mass (tons)  & Small Ring & Large Ring \\
		         \hline
			   $\nu +$D          &\cite{ku94}  &  \m35   &
			   \m2363      & \m180
			     \\
			     $\bar\nu +$D      &\cite{ku94}  &  \m35&
			     \m25779     &
			         \m1956      \\
				       $\nu + ^{16}$O    &\cite{nuO}
				       &  \m952  &  \m6054      &
				               \m734       \\
					                  $\bar\nu +^{16}$O
							  &\cite{nuO}
							  &  \m952  &
							  \m82645     &

\m9453      \\
              $\nu +^{56}$Fe    &\cite{nuFe}       &  \m250  &  \m20768 &
	                        \m1611      \\
				                       $\nu +^{208}$Pb
						       &\cite{nuPb}       &
						       \m360  &  \m103707
						           &  \m7922      \\
							           \hline
								       \end{tabular}\\[2pt]
								       {\it

Neutrino-nucleus interaction rates (events/year) at a low-energy
beta-beam facility \cite{ser04}:}
Rates on deuteron, oxygen, iron and lead are shown as examples.
The rates are obtained with $\gamma=14$
as boost of the parent ion. The neutrino-nucleus cross sections are taken
from referred references.
The detectors are located at 10 meters from the storage ring and have
cylindrical
shapes ($R$=1.5 m and $h$=4.5 m for deuteron, iron and lead,
$R$=4.5 m and $h$= 15 m for oxygen, where $R$ is the radius and $h$ is the
depth of the detector).
Their mass is indicated in the second column. Rates obtained for two
different
storage ring sizes are presented: 
the small ring has 150 m straight sections and 450 total length, while
the large ring has 2.5 km straight sections and 7 km total length.
Here 1 year = $3.2 \times 10^{7}$ s.
\end{table*}
One can see that interesting interaction rates can be obtained on one hand and
that clearly a small devoted storage ring is more appropriate for such
studies on the other hand. The physical reason is simple. Since the
emittance of the neutrino fluxes is inversely proportional to the $\gamma$ of
the ions, only the ions which decay close to the detector contribute
significantly to the number of events, while those who decay far away see the
detector under a too small opening angle. 
The complementarity  between a low-energy beta-beam and
conventional source is discussed in \cite{mcl04}.

\section{CONCLUSIONS}
The use of the beta-beam concept to produce neutrino beams in the tens of MeV
energy range is very appealing. If both electron-neutrino and
electron (anti)neutrino beams of sufficiently high intensities can be achieved,
low-energy beta-beams can offer a flexible tool, where the average neutrino
energy can be varied by varying the $\gamma$ of the ions. 
The studies realized so far indicate clearly that a small devoted storage ring
is more appropriate to obtain such beams 
in particular for performing neutrino-nucleus interaction
studies, a promising axis of research. We have particularly
discussed the interest of such measurements for a better knowledge of the
nuclear response relevant for neutrinoless double-beta decay searches.
The feasibility study of the small storage ring is now ongoing within the
Eurisol Design Study. Several issues
need to be addressed (e.g. stacking ion method, duty factor). 
The realization of low-energy beta-beams would be a {\it
proof-of-principle} that the beta-beam concept works.

\vspace{.2cm}
\noindent
We thank A. Chanc\'e, 
M. Benedikt, M. Lindroos and J. Payet for useful discussions on
the feasibility of low-energy beta-beams. The author acknowledges the 
BENE (BEams for Neutrino Experiments, CARE) fundings.

\end{document}